# A multi-agent system for managing the product lifecycle sustainability

T.MANAKITSIRISUTHI, Y.OUZROUT, A. BOURAS

Université Lumière Lyon 2,
LISEP Laboratory
160 Bd de l'université
69676 Bron Cedex

**Abstract**. The international competitive market causes the increasing of shorten product life cycle and product development process with the improvement in term of time, cost and quality while increasing the waste generation. Product life cycle sustainability can reduce waste, conserve resources, use recycling materials, design product for easy disassembly and avoid using hazardous material. This paper proposes a knowledge management architecture, based on a multi-agent system, which focuses on the "sustainability" in order to manage knowledge in each stage of the product lifecycle, and particularly in the recovery process.

The aim of this research work is to make the link between a decision-making system based on the agent's knowledge about the sustainability (environmental norms, rules…) and a PLM (Product Lifecycle Management) system. The software Agents will help the decision makers in each stage of the lifecycle and make them take into account the environmental impact of their decisions.

The proposed architecture will be illustrated in a case study.

1. INTRODUCTION

Today, increasing international competition and growing markets and the advanced technology cause the shorten product life and product development which result on the improvement of its performance in terms of time, cost and quality at the same time increasing the waste generation. The huge amounts with different kind of products have been introduced into the market to satisfy several of customers' requirements thus, the environment can be affected inevitability. The degree of environmental impacts is determined by materials and energy used in the production process and outputs generated at each stage of product's lifecycle. Managing the returned product can help in reducing the damage of environment. Environmental laws and regulations have been established to require the organization to take responsibility of return products when they come to end of their life or end of usage. Some organizations have found that returned products or used products can be the additional source of revenues by recycling materials or reusing product's components or parts after disassembling in the manufacturing process. This brings the attention of organization in managing the returned products.

In order to produce the products that are sustainable and less harmful the environment, knowledge that occurred during the production process, recovery process, and the related activities of organization including the knowledge about environmental performance should be captured, evaluated and stored.

Consideration of environmental issues and regulations during the design process, product development and particularly in recovery process can help user for decision making in each stage of lifecycle, consequently, minimizing the waste generation of product and improving the environmental performance.

This research purposes a knowledge management architecture based on multi-agents system in order to manage knowledge about the environmental management and the system will facilitate in decision making by making the link of agent's knowledge about sustainability and PLM system. The system will help decision makers in each stage of the lifecycle and make them take into account the environmental impacts of their decisions.

## 2. LITERATURE REVIEW

In the past, organizations have been asked to develop products and services which are better in quality, less expensive within timely manner. Environmental effects were overlooked during the design stage for new products and processes. Hazardous wastes were discarded inattention in the most convenient ways which damage to our environment. Concerning the environmental impact, customers are increasing their expectation that organizations are be able to handle and willing to take responsibility of their returned products (relationship between trust and reverse logistics performance) [1] and to be able to reduce the environmental impact from their activities and products.

Organizations can create their profitability from repairing, remanufacturing and recycling of returned products [2]. Since customers are conscious to environmental impact, organizations are increasing the improvement of customer satisfaction by minimizing raw materials used in the production process at the same time reducing costs of production and wastes generated into environment [3]. Therefore, reverse logistics are increasingly utilized by organizations as competitive strategy [4] [5]. Environmental legislations have been raised in order to improve the environmental performance of organization [6].

There are legislations concerned about environmental performance to encourage organizations towards sustainable product by increasing the reuse, recycle, and recovery of returned product. Many organizations have considered environmental issue as a key driver influencing to environmental performance [7] [8]. The objective of Environmental management system is to minimize the damage on the environment which cause by the organization's activities and products [9].

The influence of environmental regulations, customer awareness and social responsibility, the decision making during the design stage plays an important role to environmental performance. Thus, organization needs to design product with the improvement of effective the use of materials and energy and for better quality with less effect to environment. Moreover, in recovery process, the effective of reusing, recycling and remanufacturing of returned product can help decreasing environmental impacts and conserving resources and energy. The theoretical model of eco-oriented design [10] has presented in order for suggestion the design alternatives to designers and decision makers for balancing the use of nature resources and industry. Another research [11] shown the concept of design for sustainability, the concept will encourage organization to design products and services in dimensions of environmental, social and economic performances.

Figure 1 shows an overview of product's activities in both forward logistics and reverse logistics. Used product will be returned back to the organization with varieties reasons of returns. The inspection stage will need a lot of knowledge intensive to consider where the returned product should go to recycle, repair, resale or reuse.

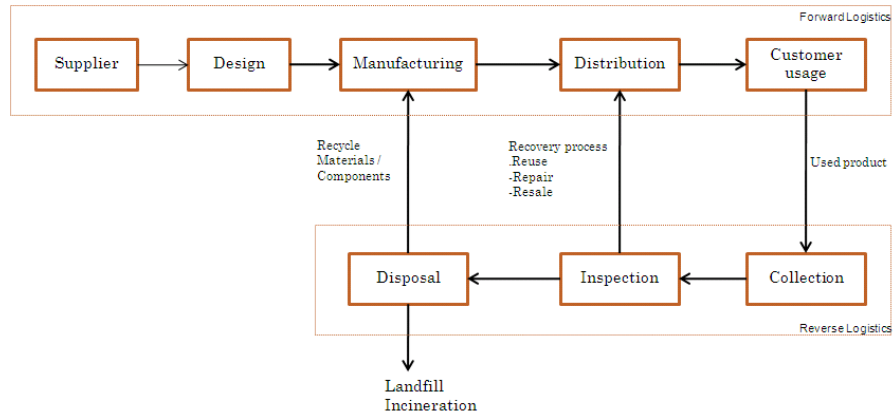

Figure1. Product lifecycle stage in forward and reverse logistics

Recovery process can help organization consume less of materials by reusing the second life of materials or parts and components from the return products which results in reduction the use of virgin of raw material and increases productivity while minimizes the environmental impacts.

Increasing awareness of need for sustainable products will result in the integration of environmental aspect into product design and product development. The proposed knowledge management architecture based on multi agent system model in order to link agents' knowledge and PLM system and support user decision making in each stage of product lifecycle related to environmental impacts.

3. KNOWLEDGE MANAGEMENT ARCHITECTURE

There are many reasons of using agents for developing knowledge management system. First, they are proactive due to user in recovery process or production process does not know or does not have time to search on product information or knowledge related to environmental performance. Therefore, agents can help user to collect and to manage those knowledge. Second, agents can make the system become more efficient because agents can learn from their own previous experiences either failures or successes [12]. Third, each agent in the system can use different technique to solve the problem such as Inspect agent uses CBR (Case-Based Reasoning) to help inspector solving problems. And the last one is agents can manage its own knowledge and interact to other agents ask for the solutions which does not exist in their knowledge base.

The architecture is composed of Inspect agent, Redesign agent, Recover agent and Disposal agent (see Figure2).

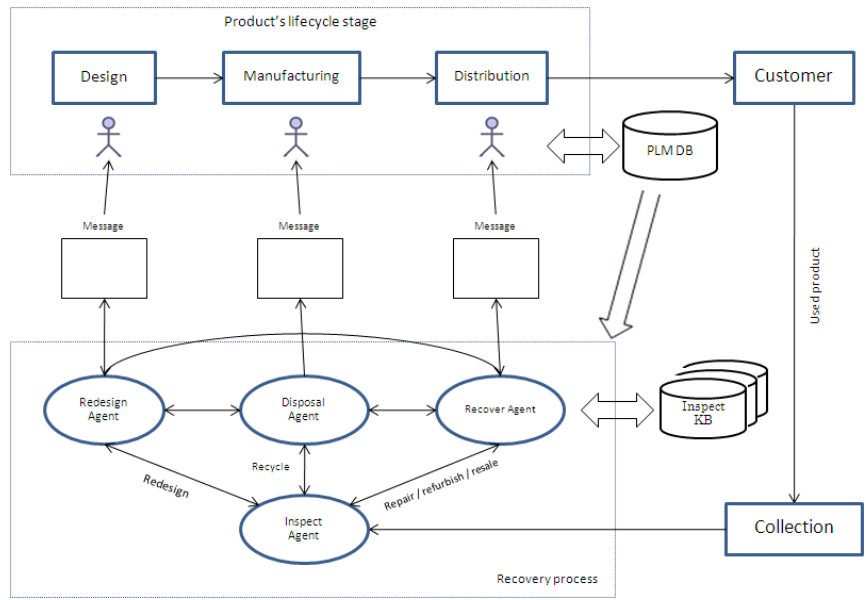

Figure2. Knowledge management system architecture based on multi-agent system

The *Inspect agent* is an agent to check the quality of returned product from the previous experience in their knowledge base. It also sends message asking the other agents if there any solutions exist (see Figure3). Then Inspect agent will determine the destination of returned product (repair, reuse, resale, recycling or redesign). For example, in case of the returned product need to be fixed or replace the damaged parts/components, Inspect agent will set the status of returned product to repair.

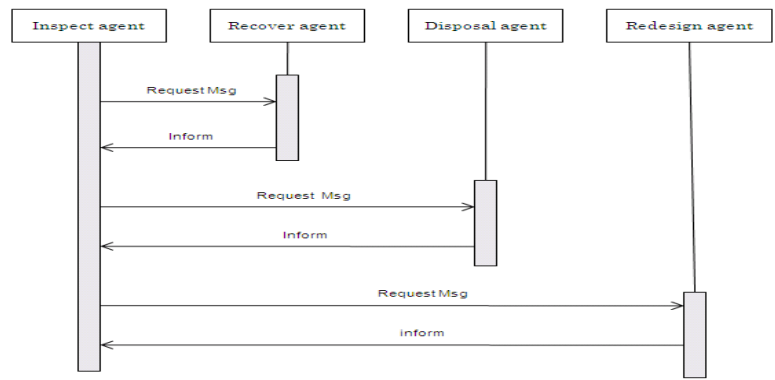

Figure3. Inspect agent sends a request to for information

The *Recover agent*, after receives the message from Inspect agent that returned product needs to be repaired, Recover agent will search all the solutions how to recover product. The solutions in Recover agent's knowledge base are the solutions concerned to the environmental performance. For example, Recover agent may send information about parts or materials which can be reused in the inspected product or inform the substitute materials.

Once an Inspect agent analyzes and finds out that the returned product is needed to redesign in order to respond customer satisfactions. *Redesign agent* will do their work by search the solution in knowledge base along with the product information in PLM system. The given solutions will be presented and correspondence with the concept of product lifecycle sustainability such as redesign the user manual to reduce the amount of paper or to extend the life of product usage in case of reason of return is customers do not know how to operate the product. In term of conserving resources and energies, Redesign agent may inform designer to increase the use of recycled materials in product, reduce weight and size of components and products.

In order to reduce environmental impact as much as possible, *Disposal agent* manages information related to reuse materials or components from the damaged product. Disposal agent can send information about the recyclable design product or inform to label material type in order to reduce the time and cost of recycling.

The aim of this research is to encourage user of making their decision in each stage of product lifecycle mainly concentrate on environmental impact of product. To develop the prototype and testing the proposed architecture based on multi agent concept, JADE (Java Agent Development

Framework) platform was chosen. JADE is an open source, implemented in java language and compliance with FIPA specifications.

4. CONCLUSION AND FUTURE WORK

This paper presents knowledge management architecture based on multi agent system to help user in each stage of product lifecycle for decision making by taking to account on environmental impacts.

The prototype is in the process of developing for testing some scenarios for product lifecycle sustainability.

REFERENCE

1. Daugherty, P.J., Richey, R.G., Hudgens, B.J. and Autry, C.W., "Reverse logistics in the automobile aftermarket industry", International Journal of Logistics Management, 2003, Vol. 14, No. 1, pp. 49-61.

2. Giuntini, Ron and Andel, Tom," Master the six R's of reverse logistics" Transportation and Distribution, March 1995.

3. Ayres, R.U., Ferrer, G., Van Leynseele, T., 1997. "Eco-efficiency, asset recovery and remanufacturing". Eur. Manag. J. 15 (5), 557–574.

4. Edward J. Marien, "Reverse Logistics as competitive strategy", Supply Chain Management Review, 1998.

5. Stock, J.R., Speh, T.W., Shear, H.W., 2002. Many happy (product) returns. Harvard Business Review 80 (7), 16-17

6. Bloemhof J, van Nunen J. "Integration of environmental management and SCM". ERIM, Report Series Research in Management 2005; ERS-2005-030-LIS.

7. H.-J. Bullinger, J. von Steinaecker, A. Weller "Concepts and methods for a production integrated environmental protection", International Journal of Product Economics, 60-61 (1999) 35-42

8. Shane J. Schvaneveldt, "Environmental performance of products: Benchmarks and tools for measuring improvement", Benchmarking: An International Journal, 2003, vol.10, issue 2, p 137-152.

9. http://www.iso.org/iso/iso_catalogue/management_standards/iso_9000_iso_14000/iso_14000_essentials.htm

10. Hernane Borges de Barros Pereira and Paulo Fernando de Almeida Souza, "Design for sustainability: Method in search for a better harmony between industry and nature".

11. Marcel Crul, Jan Carel Diehl, "Design for Sustainability (D4S): Manual and Tools for Developing Countries", 2008, 7th Annual ASEE Global Colloquium on Engineering Education.

12. P. Maes, "Agents that reduce work and information overload", Communications of the ACM, 37(7), (1994), p. 31-40.